# The Design and Evaluation of the Cloud-based Learning Components with the Use of the Systems of Computer Mathematics

Mariya Shyshkina[1][0000-0001-5569-2700], Uliana Kohut[2][0000-0002-2861-2274],
Maiia Popel[1][0000-0002-8087-962X]

[1]Institute of Information Technologies and Learning Tools of NAES of Ukraine,
9 M. Berlynskoho St., Kyiv, Ukraine
`{shyshkina, popel}@iitlt.gov.ua`
[2]Drohobych Ivan Franko State Pedagogical University, 24 I. Franko Str., Drogobych
`ulyana_kogut@mail.ru`

**Abstract.** In the article the problems of the systems of computer mathematics use as a tool for the students learning and research activities support are investigated. The promising ways of providing access to the mathematical software in the university learning and research environment are considered. The special aspects of pedagogical applications of these systems to support operations research study in the process of bachelors of informatics training are defined. The design and evaluation of the cloud-based learning components with the use of the systems of computer mathematics (on the example of Maxima system) as enchasing the investigative approach to learning of engineering and mathematics disciplines and increasing the pedagogical outcomes is justified. The set of psychological and pedagogical and also technological criteria of evaluation is substantiated. The results of pedagogical experiment are provided. The analysis and evaluation of existing experience of mathematical software use both in local and cloud-based settings is proposed.

**Keywords:** Cloud computing, systems of computer mathematics, learning tools, mathematical disciplines, learning environment, educational university.

## 1   Introduction

### 1.1   Research objectives

In the modern information-educational environment there are new models of learning and research activity organization that are based on innovative technological solutions. The question of this environment facilities and services organization to enhance the pedagogical effect of modern ICT use comes to the fore. To attain the increase of learning outcomes and also the improvement of the students' research activity the innovative educational approaches are in demand.

A separate set of problems concerns to the application of software packages for the implementation of various mathematical operations, actions and calculations, these are the so-called Systems of Computer Mathematics (SCM), including Maple Net, MATLAB web-server, WebMathematica, Calculation Laboratory and others [12; 14]. These systems are the most common types of mathematical software, being a part of the modern learning environment of educational institutions [3, 4, 5, 7, 8]. The problems emerge when searching for promising methods and models of these systems use to enhance the pedagogical outcomes and provide the investigative approach to learning of engineering and mathematics disciplines.

**The aim of the article** is justification of the cloud-based learning components design with the use of the systems of computer mathematics (on the example of Maxima system) as enchasing the investigative approach to learning of engineering and mathematics disciplines and increasing the pedagogical outcomes.

### 1.2 The Problem Statement

Nowadays, SCM make a significant impact on the content and forms of learning mathematics and informatics disciplines in higher educational institutions.

A separate set of problems relates to the use of mathematical software tools to enhance the investigative approach to learning. There are two factors in this respect that may significantly influence the investigative activity of students. Firstly SCM bring the possibility to address the basic notions of mathematics and informatics on the research level. Due to this the concepts of soft computation, discrete mathematics and others that are mainly computer oriented are included into the learning content. Secondly SCM being the tool for computer modeling as the general method of investigation that is the fundamental base of all mathematics and computer science disciplines become the instrument of research.

There is a significant demand in expansion of access to research activities tools while learning informatics and mathematics disciplines in educational universities as well as modernization of the learning environment with the use of current ICT tools, especially the cloud-based ones.

The progress in the area has provided new insights into the problems of educational learning environment development, bringing new models and approaches. These tools make the great impact on the learning data processing changing the content, methods and organizational forms of learning, lifting the restrictions or significantly improving the access for all participants [10].

So, the modeling and analysis of the learning components design and deployment and available learning experience of its use in view of the current tendencies of modern advance of the cloud-based mathematical software have come to the fore.

### 1.3 The Research Methods

The research method involved analyzing the current research (including the domestic and foreign experience of the cloud-based learning services and mathematical software use in educational institutions in Ukraine and abroad), evaluation of existing approaches to software delivery, their advantages and disadvantages; comparison of promising ways of popular mathematical software implementation "in the cloud", examining of the models and approaches, technological solutions and psychological and pedagogical assumptions about better ways of introducing innovative technologies into the learning process. The cloud-based component with the use of Maxima system was designed and elaborated within the study undertaken in 2012-2014 in the Institute of Information Technologies and Learning Tools of NAES of Ukraine devoted to the use of the SCM for the informatics bachelors training (U. Kohut). The special indicators to reveal ICT competence of educational personnel trained within the cloud-based learning environment and also the learning components quality evaluation indicators were elaborated within the research work devoted to the university cloud-based learning and research environment formation and development held in 2012-2014 in the Institute of Information Technologies and Learning Tools of NAES of Ukraine (M. Shyshkina). To measure the efficiency of the proposed approach the pedagogical experiment was undertaken in Drohobych Ivan Franko State Pedagogical University. The expert quality evaluation of the cloud-based components elaborated in the study was implemented. The approach and methodology were grounded within the research work "Methodology of the cloud-based learning environment of educational institution formation" that was held in the Institute of Information Technologies and Learning Tools of NAES of Ukraine in 2015-2017, Registration number 0115U002231 (coordinated by M. Shyshkina).

## 2  The State of the Art

The analysis of the domestic and international experience of ICT use of in the process of informatics disciplines learning testifies that such class of ICT-based learning tools as the systems of computer mathematics(SCM) constantly attracts an attention of researchers [3; 4; 5; 8]. These systems, that are complex, multifunctional, powerful enough and at the same time simple in the use, become irreplaceable in maintenance of various processes of numerical accounts, patterns visualization, realization of symbol operations, algorithms and procedures [7; 8]. SCM is the environment for design and use of learning tools and components for informatics and mathematics disciplines, forming innovative pedagogical technologies.

In the recent years, the informatics disciplines learning tools and technologies have been actively developed with the use of the cloud computing approach [3, 4, 8]. This conception significantly changes the existing views on the organization of access and integration of applications, so there is a possibility to manage larger ICT infrastructures that allow to create and use both individual and collective "clouds" in a cloud-oriented educational space [1; 7].

Localization of such tools as SCM "in the cloud" is the perspective trend of their development, when there are more possibilities for adapting the learning environment to educational demands, individual needs and goals of the learners. There is expansion of a "spectrum" of research activities due to both fundamentalization of informatics disciplines teaching content and expansion of access to research activities tools. In this regard there is a need to consider the issues of theoretical and methodical grounding of the SCM-based learning components design, revealing advantages and disadvantages of different approaches to their deployment and implementation.

The use of SCM Maxima in the process of operations research study aims at the forming of students' ICT-competences due to: the acquaintance with functional characteristics of SCM Maxima; developing skills of mathematical research of the applied tasks, in particular the construction of mathematical models; mastering programming in the SCM Maxima environment; obtaining the necessary knowledge base for studying other math and informatics disciplines; increasing the level of informatics acquirement by means of the extensive use of SCM and cloud based systems in the educational process and research work.

Methodical peculiarities of teaching optimization methods and operations research using WEB-SCM are analyzed in the work of Trius Y. V. [6]. The graphical interface of SCM Maxima for modeling animations is described in detail in the work of Bugaets N. O. and examples of creating the animation evident models and their use for development of educational-research abilities are given [2]. The problems of the right choice of SCM to support learning and research activity and elaboration of the most advisable methods of its use for math and computer science discipline so as to enhance the investigative activity of the students remain crucial in the area.

## 3    The Research Results

The choice of SCM to support the investigative approach to learning depends on the input data and results to be obtained. For example, the analytical model of the investigated phenomenon or object is more interesting for a physicist-theorist, so it is better to use the packages such as Mathematica, Maple and Maxima. Physicists-experimenters would rather use the MATLAB system for large data sets processing [8, p. 138].

Special attention should be paid to Maxima system, as it is easy in learning, in solving the problems does not yield to such systems as Maple and Mathematica and is freely distributable. It is equipped with a menu system that allows to perform symbol conversions, solve equations, compute limits, derivatives, integrals and the like, without mastering the language for description of the commands to perform these actions. Therefore, Maxima system can be used for informatics and mathematics disciplines learning even in the first course of educational university [8]. Maxima system introduction will not cause any difficulties for students in solving tasks of mathematical analysis and linear algebra – the students are required only to select a menu item and enter the expression. However, for programming in Maxima system one needs knowledge of language and syntax, as well as certain commands [8, p.138].

The goal of SCM use in the process of informatics bachelors training in educational university is the formation of the ability for successful application of the information technologies in their professional activities, enhance the creative approach to solving non-standard problems, deep mastering the fundamentals of the disciplines. For this purpose the methodology of SCM using in the process of operations research teaching was developed, aimed at (I) the formation of the professional competences of pre-service informatics teachers that will give an opportunity in the future to adapt oneself to the requirements of informative society; (II) the development of the creative approach to solving non-standard tasks; and (III) the formation of mathematical skills needed for analyzing, modeling and solving theoretical and practical problems with application of SCM [8]. The use of this technique was a subject of the experimental studies with application of both local and cloud-oriented implementation of SCM Maxima [8].

The field of operations research study requires special attention as it combines both the fundamental concepts and principles of different mathematics and informatics disciplines and applied models and algorithms for their application.

Due to the introduction of SCM Maxima into the operations research teaching process the opportunity to focus students on key concepts, principles, approaches, releasing time and efforts that are spent on the software establishment, maintenance, and even greatly to mitigate the real spatial and temporal boundaries of the implementation of access to necessary electronic resources occurred. This approach develops interdisciplinary links, assists the deep study of material, and extends possibilities of independent research, combination of theory and practice, knowledge integration concerning the various departments and levels of computer education [7; 8].

The use of the cloud-based tools of SCM design is a significant factor in the expansion of access to them in the process of teaching and research activities in the field of informatics and mathematics. If research activity was provided only in specially created situations in the case of application of a local version of the tool, in case of the cloud-based version more attention can be paid to the independent work, and research activity is extended outside the classroom time [8].

For this purpose, the technology of "virtual desktop" was applied, where the data storage and processing were maintained in the data center. Also, for a user, the work with cloud applications, appealed via the Internet browser, does not differ from the work with software installed on a desktop of the user's personal computer [8].

The use of software that is installed on the student's virtual desktop (I) does not require spending learning time on installing and updating, (II) the conditions for more differentiated approach to learning are created, and (III) provides the opportunity to focus on the basics of the teaching material [8].

The necessity to use SCM in the educational process is also caused by the fact that working with them provides students with the real opportunity to acquire skills to solve practical problems using the conventional scheme: setting of the problem →defining modeling goals → mathematical model development → election of mathematical method and algorithm of problem solution → implementation of mathematical model using SCM → calculations → analysis of the results obtained and their interpretation → making the decision.

A large number of practical problems are studied within the discipline "Operations research", which are easy to interpret as optimization problems on graphs. The examples of such tasks are (i) searching for the shortest route between two settlements, (ii) determination of the maximal admission characteristics of the oil pipeline, and (iii) scheduling the execution of the project works etc.

When solving optimization problems on graphs the interdisciplinary relationships of informatics, mathematics, economics and other disciplines are realized that contributes to the intellectual development of students on the basis of forming ideas about the integrity of vision of the world, ensures the formation of skills and not only declarative but also procedural knowledge. The graph theory problems solution develops the students ability to represent the problem in the graph theory language, and then to interpret the solution in terms of the original problem.

The possibilities of using Maxima system to solve optimization problems on graphs are wide enough. A student, using SCM Maxima, solves the problem set before him, and thus he doesn't have the psychological barrier in application of mathematical apparatus, and besides he realizes also, what material is necessary to be repeated (or to be learnt). The solution of problems of applied nature (including, in particular, optimization problems on graphs) using a SCM provides the possibility of formation of the professional competences. The interest is also the research of optimization theory problems, in particular the implementation of the numerical methods, both conditional and unconditional optimization using SCM Maxima.

Studying the section "Models of the dynamic programming" the students are offered to solve the problems which demand using Maxima commands and functions or creating their own procedures and functions. This in turn contributes to the improvement of programming skills. For example, when solving the problem of dynamic programming about a backpack, students perform research, creative work, and its routine is completed using the computer.

The main stages of the solution of such problems are the problem setting (providing of the objective function, optimality criterion, limitations, and accuracy of the solution) and analysis of the obtained results. The students get the system approach basis in solving problems, and they see the relationship of the content of various academic disciplines studies.

Summarizing the consideration of the course "Operations research", it should be noted that a wide set of tools for computer support of analytical, computing and graphical operations make the system of computer mathematics to be one of the main tools in the professional activities of mathematicians and programmers. The studies using Maxima system combine algebraic and computing methods. In this sense, SCM is the combining link between mathematics and computer science, where the research focus both on the development of algorithms for symbolic computation and data processing using computer and the creation of the programs to implement these algorithms.

### 3.1 The Design of the Cloud-based Learning Component with the Use of the Maxima System

To research the hybrid service model of learning software access, especially for the mathematical software delivery a joint investigation was undertaken in 2013–2014 at the Institute of Information Technologies and Learning Tools of the NAES of Ukraine and Drohobych State Pedagogical University named after I. Franko. At the pedagogi-cal experiment the cloud-based learning component with Maxima system was designed and used for the operations research study [7].

In this case, the implementation of software access due to the hybrid cloud deployment was organised [7].

The configuration of the virtual hybrid cloud used in the pedagogical experiment was described in [7]. The model contains a virtual corporate (private) subnet and a public subnet. The public subnet can be accessed by a user through the remote desktop protocol (RDP). In this case, a user (student) refers to certain electronic resources and a computing capacity set on a virtual machine of the cloud server from any device, anywhere and at any time, using the Internet connection.

The advantage of the proposed model is that, in a learning process, it is necessary to use both corporate and public learning resources for special purposes. In particular, the corporate cloud contains limited access software; this may be due to the copyright being owned by an author, or the use of licensed software products, personal data and other information of corporate use. In addition, there is a considerable saving of computational resources, as the software used in the distributed mode does not require direct Internet access for each student. At the same time, there is a possibility of placing some public resources on a virtual server so the learner can access them via the Internet and use the server with the powerful processing capabilities in any place and at any time. These resources are in the public cloud and can be supplied as needed [8].

Within the experimental study the Maxima system installed on a virtual server running Ubuntu 10.04 (Lucid Lynks), was implemented. In the repository of this operational system is a version of Maxima based on the editor Emacs, which was installed on a student's virtual desktop [8].

To create a session (to insert an item Maxima) you should choose the menu option Insert – Session – Maxima. There is an active input line to input Maxima commands (Fig.1).

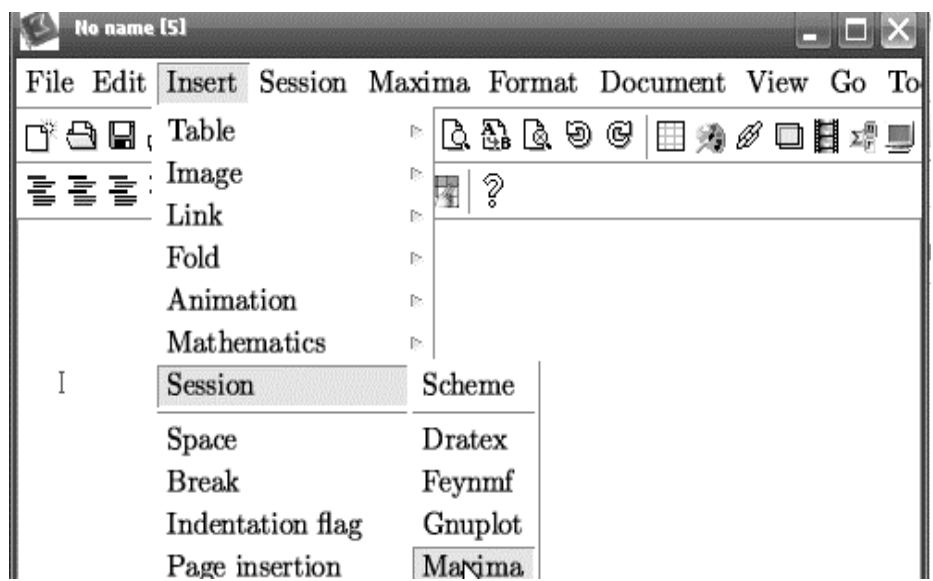

**Fig. 1.** Insert the Maxima object into a document

The promising experience of successful use of the Maxima cloud-based component was achieved in the Graph Theory learning [8]. Maxima has a rich set of features on the design and elaboration of relevant objects of this theory. Some examples of its use are presented at [8].

### 3.2   The Results of the Pedagogical Experiment

During 2010-2014 the experimental research was being conducted. During the experiment SCM MAXIMA was implemented in the process of operations research teaching concerning the students of the Institute of Physics, Mathematics, Economics and Information Technology of the Drohobych Ivan Franko State Pedagogical University (education and qualification level "Bachelor", area of knowledge – 0403 "System sciences and cybernetics", areas of training – 6.040302 " Informatics"). In the experiment the specially developed learning method of operations research teaching using Maxima system was tested. At the formative stage of the experiment there was 240 students participated. The experiment confirmed the research hypothesis concerning the increase of the level of professional competences development in the process of study due to the use of the proposed learning technique [8, 10]. It was also showed that by means of the cloud technology the students can get better access to the research activity tools and facilities.

In the experiment both the local version of the system installed on the student computer desktop and the cloud-based version that was posted on the virtual desktop were involved.

The results of formative stage of pedagogical experiment in the control and experimental groups and comparative histogram distribution of the students learning achievements due to the results of the final exam by discipline "Operations Research" is shown in Fig. 2.

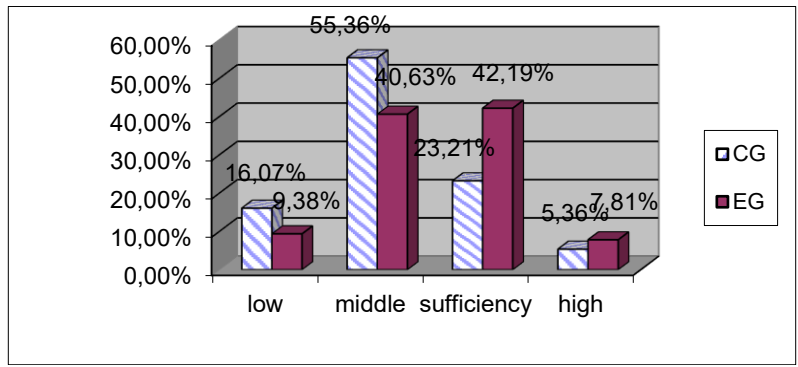

**Fig. 2.** A comparison of the students learning achievements by the results of the final control exam by the course "Operations Research" after the forming stage of the experiment

The level of students' knowledge on the course "Operations research" as well as professional disciplines was checked according to the results of complex state examination to justify the influence of methodology of SCM using as operations research teaching tools on the increase of the level of some components of professional competence.

Null hypothesis $H_0$: distribution of students scores on operations research in the control ($n_1 = 56$) and experimental samples ($n_2 = 64$) after the formative forming stage of the experiment do not differ ($i = 0, 1, …, 6$).

$Q_{1i}$ – number of participants in the control group who scored i points;

$Q_{2i}$ – number of participants in the experimental group who scored i points.

Alternative hypothesis $H_1$: distribution of students scores on operations research in the control ($n_1 = 56$) and experimental samples ($n_2 = 64$) after the formative forming stage of the experiment differ ($i = 0, 1, …, 6$).

The calculation results of statistics of these samples are given in table 2.

**Table 1.** Calculation of $\chi^2$ for the control and experimental groups after the formative experiment on the course "Operations research"

| *I* | $Q_{1i}$ | $Q_{2i}$ | $S_{12i}$ |
|---|---|---|---|
| 0 (*F*) | 0 | 0 | 0 |
| 1 (*FX*) | 18 | 12 | 30720,00 |
| 2 (*E*) | 50 | 22 | 215168,00 |
| 3 (*D*) | 12 | 30 | 79213,71 |
| 4 (*C*) | 14 | 34 | 84672,00 |
| 5 (*B*) | 12 | 20 | 15488,00 |
| 6 (*A*) | 6 | 10 | 7744,00 |
| **T** | | | **30,20408** |

The calculation of $\chi^2$ criterion for the experimental and control samples after conducting the formative stage of the experiment showed that Texp > Tcritical (30,20408> 11,07). This is the reason for rejecting the null hypothesis.

The acceptance of alternative hypothesis suggests that these samples have statistically significant differences, i.e., the experimental method is more effective than the traditional one.

Considering that in the experimental groups the training of students was performed according to the developed technique, it can be assumed that this contributed to the achievement of better results. Therefore, it is possible to confirm the hypotheses of the research.

Summarizing, we conclude that the pedagogical experiment confirmed the hypothesis of the study. Analysis of the results indicates the increase of the learning outcomes level due to the enhancing the investigative component in the learning process using the developed method.

The special aspect of the study was the learning method application using the cloud version of the Maxima system that was posted on a virtual desktop. In the first case study (with the local version), this tool was applied only in special training situations. In the second case study (the cloud version) the students' research activity with the system extended beyond the classroom time.

The cloud-based learning component used in the experiment has undergone a quality estimation. The method of learning resources quality estimation developed in the joint laboratory of educational quality management with the use of ICT [7] was used and adapted for this study. The 20 experts were specially selected as having experience in teaching professional disciplines focused on the use of ICT and being involved in the evaluation process. The experts evaluated the electronic resource by two groups of parameters. The first group contained 7 technological parameters: ease of access; the clarity of the interface; sustainability; support of collaborative work, ease of integration; mobility; and usefulness. The second group contained 9 psychological and pedagogical parameters: the scientific clarity; accessibility; fostering the intellectual development; problem orientation; personalization; adaptability; methodical usefulness; professional orientation; and feedback connection. The results of the quality parameters valorisation and the experts' concordance research are described in [9].

The problem was: is it reasonable and feasible to arrange the environment in a proposed way? For this purpose there were two questionnaires proposed to expert concerning two groups of parameters. The 20 experts estimated 16 parameters (there were 7 technological and 9 psychological and pedagogical among them). A four-point scale (0 (no), 1 (low), 2 (good), 3 (excellent)) was used for the questions.

The resulting average value was calculated for every parameter among the technological ones : "Ease of access" = 2.1, "Interface clarity" = 2.4, "Responsiveness" =

2.1, "Sustainability" = 2.56, "Support of Collaborative work" = 2.0, "Ease of Integration" = 2.0, "Usefulness" = 2.8, the total value was 2.3 [9].

The resulting average values for every psychological and pedagogical parameter was calculated as: "Scientific clarity" = 2.6, "Accessibility" = 2.7, "Fostering the intellectual development" = 2.5, "Problem orientation" = 2.8, "Personalization" = 2.8, "Adaptability" = 2.6, "Methodical usefulness" = 2.81, "Professional orientation" = 2,75, "Feedback connection" = 2,75. The total value was 2.71 [9].

The resulted average criterion of EER quality K=2,59. This characterises the resource quality as sufficient for further implementation and use [8, 9].

We can see that the results of the cloud-based component evaluation by the set of technological and also psychological and pedagogical indicators reveal the usefulness of this component to support the investigative approach to learning. The highest scores of the parameters values are "Scientific clarity" = 2.6, "Accessibility" = 2.7, "Fostering the intellectual development" = 2.5, "Problem orientation" = 2.8, "Personalization" = 2.8, "Adaptability" = 2.6, "Methodical usefulness" = 2.81, "Professional orientation" = 2,75, "Feedback connection" = 2,75. Just these kinds of indicators are the most important and "responsible" for the investigative activity of the learner. This fact also supports the hypothesis that the introduction of SCM into the learning process in particular within the cloud-based settings really extends the boundaries of the students research activities expanding it into the broader context.

The advantage of the approach is the possibility to compare the different ways to implement resources regarding the learning infrastructure. Future research in this area should consider different types of resources and environments.

## 4      Conclusions and Discussion

The results of the study indicate certain movement in the development of new ways to create and use of the software for educational purposes.

The use of mathematical packages to support the investigative approach to learning involves (I) understanding of the problems of the learning domain for proper use of SCM; (II) understanding the methodology of developing the algorithm from the mathematical statements and formation of the ability to apply this methodology; and (III) the ability to carry out the estimation of the algorithm `at run-time and memory requirements. In this case SCM is to provide the tools for modelling and research of the domain objects in the learning process, to make experiments and approve the results.

The introduction and design of the cloud-based learning components into the process of math and computer science training contributes to the growth of access to the best examples of electronic resources and services to support the research activities within the learning process. The use of these technologies adds and provides an opportunity to explore and develop investigative approach to learning, which in turn leads to the development of new strategies and methodology of teaching of mathematics disciplines in educational universities. It brings the possibility to expand the investigative activity of students beyond the classroom, to provide the tools for modeling and research of the domain objects in the learning process, widening the spectrum of research activity due to the content fundamentality and interdisciplinary links establishment.